\documentclass[journal=jctcce]{achemso}
\setkeys{acs}{articletitle=true}

\pdfoutput=1
\usepackage[utf8]{inputenc}
\usepackage[T1]{fontenc}
\usepackage{lmodern}
\usepackage{microtype}

\usepackage{amsmath}
\usepackage{graphicx}
\usepackage{hyperref}
\usepackage{xcolor}

\definecolor{one}{HTML}{CC0000}

\hypersetup{
  colorlinks=true,
  citecolor=one,
  linkcolor=one,
  urlcolor=one
}

\newcommand*{\ev}[1]{\langle #1 \rangle}
\newcommand*{\bra}[1]{\langle #1 \rvert}
\newcommand*{\ket}[1]{\lvert #1 \rangle}

\author{Miroslav Urbanek}

\email{urbanek@lbl.gov}

\author{Daan Camps}

\author{Roel Van Beeumen}

\author{Wibe A. de Jong}

\email{wadejong@lbl.gov}

\affiliation{Computational Research Division, Lawrence Berkeley
  National Laboratory, Berkeley, CA 94720, USA}

\title{Chemistry on quantum computers with virtual quantum subspace
  expansion}

\begin{document}

\begin{abstract}

Simulating chemical systems on quantum computers has been limited to a
few electrons in a minimal basis. We demonstrate experimentally that
the virtual quantum subspace
expansion~[\href{https://link.aps.org/doi/10.1103/PhysRevX.10.011004}{Phys. Rev. X~\textbf{10},
    011004 (2020)}] can achieve full basis accuracy for hydrogen and
lithium dimers, comparable to simulations requiring twenty or more
qubits. We developed an approach to minimize the impact of
experimental noise on the stability of the generalized eigenvalue
problem, a crucial component of the quantum algorithm. In addition, we
were able to obtain an accurate potential energy curve for the
nitrogen dimer in a quantum simulation on a classical computer.

\end{abstract}

It is expected that major applications of quantum computing will be in
quantum chemistry~\cite{Reiher:2017, Cao:2019, McArdle:2020}. Typical
chemistry problems are to find the ground-state energy of a molecule,
its excited states, or to extract reduced density matrices that can be
used to compute various molecular properties relevant to science and
industry. Traditional quantum algorithms, for example quantum phase
estimation (QPE)~\cite{AspuruGuzik:2005}, require circuit depths that
are beyond the abilities of currently available quantum computers. A
lot of effort has been invested into alternative approaches, most
notably into the variational quantum eigensolver (VQE)
algorithm~\cite{Peruzzo:2014, McClean:2016}. VQE is an iterative and
hybrid quantum-classical method, where one creates a parametrized
trial wave function on a quantum computer, measures observables that
correspond to Hamiltonian terms of the studied molecule, estimates the
electronic energy, and optimizes the set of wave-function parameters
for the next iteration. It has been demonstrated that VQE can find the
ground-state energy of small molecules~\cite{Peruzzo:2014,
  OMalley:2016, Kandala:2017, Shen:2017, Colless:2018,
  Dumitrescu:2018, Hempel:2018, Ganzhorn:2019, Kokail:2019}.

While the ground-state energy is an important property of a molecular
configuration, the energy of excited states is even more
important. The quantum subspace expansion (QSE)
algorithm~\cite{McClean:2017} can extract the energy of excited states
described by the occupied and unoccupied orbitals in the active space
and was also found to improve the ground-state energy
estimate~\cite{BonetMonroig:2018, Colless:2018}. Using this method,
one first creates the ground-state wave function on a quantum computer
and then performs extra measurements to analyze its single-particle or
double-particle excitations. QSE does not require additional qubits or
deeper circuits than VQE. Since this pioneering work, other extensions
of VQE that target excited states within the active orbital space have
been proposed and demonstrated~\cite{Santagati:2018, Ollitrault:2019,
  Parrish:2019}.

In most of the experimental realizations of molecular calculations on
quantum computers, only small numbers of orbitals that constitute a
subset of the basis of a many-electron wave function have been
considered. While this is sometimes a reasonable approximation, not
including additional virtual orbitals described by the chosen basis
set limits the accuracy of molecular properties, for example reaction
energetics and barriers, obtained with the algorithms discussed
above. The number of orbitals accounted for in the simulation drives
the number of qubits required and can quickly exceed those available
on near-term quantum hardware. Circuit depths increase significantly
with the number of orbitals used in the simulation as well. The
virtual quantum subspace expansion (VQSE) algorithm, proposed in
Ref.~\cite{Takeshita:2020}, is an extension of QSE that includes
excitations into virtual orbitals outside the chosen active space
without the need for additional quantum resources. The authors
analyzed VQSE in a numerical study and showed that it can improve
accuracy of chemistry calculations. The algorithm assumes that strong
correlations can be described by a subset of orbitals. Similarly to
QSE, one creates the ground-state wave function using this subset on
the quantum computer. VQSE then requires performing additional
measurements to account for the so-called virtual orbitals that were
not explicitly included in the determination of the ground-state
energy, allowing one to estimate energy levels more accurately. VQSE
scales polynomially with the size of the virtual orbital space. In
principle, the other excited state approaches mentioned above could be
adapted in a similar fashion to include additional virtual orbitals.

In this letter, we implement and execute the VQSE algorithm on a real
quantum computer available on the IBM Q Hub and calculate the
lowest-energy potential energy curves of the hydrogen, lithium, and
nitrogen dimers. We find that the noise from the quantum computer
significantly impacts the generalized eigenvalue problem that needs to
be solved classically and develop and demonstrate an approach to
overcome this issue. Our results show that VQSE works very well in
experiments and even on imperfect and noisy devices.

VQSE proceeds in three steps. Firstly, we use a quantum computer to
find the ground state in the active space. We do this using the VQE
algorithm. Next, we measure expectation values of additional
observables for the ground state. Finally, we use the measured
expectation values to calculate corrections originating from the
virtual space.

The electronic Hamiltonian is discretized into a finite set of
orbitals. We divide the orbitals into core, active, and virtual
orbitals. Core orbitals are considered frozen, i.e., these orbitals
are doubly occupied and electrons in them are never excited into other
orbitals. Since the Hamiltonian with frozen core orbitals can be
transformed into a Hamiltonian with only active and virtual orbitals,
we ignore the core space from now on. Active orbitals are the crucial
part of the system because electrons in these orbitals are typically
strongly correlated. Virtual orbitals give rise to corrections for
quantities found by taking only the active orbitals into account.

We first create a set of expansions operators~\cite{Takeshita:2020},
\begin{equation}
  \label{operators}
  S = \{a_i^\dagger a_p, a_\mu^\dagger a_q a_\nu^\dagger a_r | i \in
  \mathcal{A} \cup \mathcal{V}; p, q, r \in \mathcal{A}; \mu, \nu \in
  \mathcal{V}\},
\end{equation}
where $a^\dagger_i$ ($a_i$) is the creation (annihilation) operator
for an electron in spin-orbital $i$, and $\mathcal{A}$ and
$\mathcal{V}$ are sets of active-space and virtual-space spin-orbital
indices, respectively. Let $\ket{\Psi}$ be the ground-state wave
function of a Hamiltonian $A$ restricted to the active space. States
$O_i \ket{\Psi}$, where $O_i \in S$, are single or double excitations
of $\ket{\Psi}$. We next create a matrix $A$ representing the
unrestricted Hamiltonian in the expanded set of wave functions. $A$ is
given by its elements
\begin{equation}
  A_{ij} = \bra{\Psi} O_i H O_j \ket{\Psi},
\end{equation}
where $O_i, O_j \in S$. States $O_i \ket{\Psi}$ are generally not
orthonormal. To find the energy spectrum in the expanded set, it is
thus necessary to solve a generalized eigenvalue problem
\begin{equation}
  \label{eigenvalues}
  AC = BCE,
\end{equation}
where the overlap matrix $B$ is given by its elements
\begin{equation}
  B_{ij} = \bra{\Psi} O_i O_j \ket{\Psi},
\end{equation}
$C$ is a matrix of eigenvectors, and $E$ is a diagonal matrix of
eigenvalues.

The operators $O_i O_j$ and $O_i H O_j$ can be transformed using the
Jordan--Wigner~\cite{Jordan:1928} or a similar transformation to qubit
operators. If $\ket{\Psi}$ is the ground-state wave function created
on a quantum computer, the expectation values $\bra{\Psi} O_i O_j
\ket{\Psi}$ and $\bra{\Psi} O_i H O_j \ket{\Psi}$ correspond to
expectation values of strings of Pauli operators acting on
qubits~\cite{Takeshita:2020}.

We start with the simplest cases, finding the ground-state energy of
the hydrogen and lithium dimers. We represent their electronic wave
functions in the cc-pVDZ basis~\cite{Dunning:1989}. The full molecular
Hamiltonian of each molecule contains thousands of terms. We divide
the Hilbert space of $\mathrm{H}_2$ into an active space with two
orbitals and a virtual space with eight orbitals. Similarly, we divide
the space of $\mathrm{Li}_2$ into a core space with two orbitals, an
active space with two orbitals, and a virtual space with six orbitals.

Since our active spaces contain only two orbitals, they can be mapped
to four qubits. We are targeting ground states with two electrons and
zero total spin. This subspace contains only four basis states and can
be mapped to two qubits~\cite{OMalley:2016, Colless:2018,
  Hempel:2018}. We therefore reduce our four-qubit active-space
Hamiltonian to a two-qubit Hamiltonian to further simplify the
problem. We do this by projecting the Hamiltonian onto a subspace of
wave functions that describe two electrons with opposite spins. The
basis of this subspace is given by four-qubit states $\ket{0011}$,
$\ket{0110}$, $\ket{1001}$, and $\ket{1100}$, where even and odd
qubits represent spin-up and spin-down electrons, respectively. These
four states are then mapped to the basis states of our two qubits. The
projected Hamiltonian is given by
\begin{equation}
  \label{hamiltonian}
  H = g_1 I + g_2 Z_1 + g_3 Z_2 + g_4 Z_1 Z_2 + g_5 Y_1 Y_2,
\end{equation}
where coefficients $g_i$ are calculated numerically. A transformation
of the Hamiltonian to a smaller number of qubits is in principle
possible also for larger systems. However, it is generally not
scalable.

We use the VQE algorithm with the unitary coupled-clusters (UCC)
ansatz~\cite{Bartlett:1989, Taube:2006, Peruzzo:2014, OMalley:2016,
  Hempel:2018} to find the ground state in the active space reduced to
two qubits. The ansatz is given by
\begin{equation}
  \label{ansatz}
  \ket{\psi(\theta)} = e^{-i \theta Y_1 X_2 / 2} \ket{\Phi},
\end{equation}
where $Y_1$ and $X_2$ are Pauli Y and X matrices acting on the first
and second qubit, respectively, and $\ket{\Phi} = \ket{00}$ is the
Hartree--Fock wave function. The wave function energy is given by
\begin{equation}
  \label{energy}
  E(\theta) = g_1 + g_2 \ev{Z_1}_\theta + g_3 \ev{Z_2}_\theta + g_4
  \ev{Z_1 Z_2}_\theta + g_5 \ev{Y_1 Y_2}_\theta,
\end{equation}
where $\ev{O}_\theta = \bra{\psi(\theta)} O \ket{\psi(\theta)}$. The
ground state can be found by minimizing $E(\theta)$. Since
ansatz~\eqref{ansatz} has one parameter only, we sweep the full domain
of $\theta$ and perform the minimization during post-processing. We
also measure other expectation values to estimate the elements of $A$
and $B$.

Energy levels of a molecule are the eigenvalues of the generalized
eigenvalue problem~\eqref{eigenvalues}. Elements of the $A$ and $B$
matrices are linear combinations of measured expectation values. They
are noisy due to imperfections of real quantum computers and also due
to shot noise. The noise perturbs the eigenvalue problem and results
in nonphysical modes. As a result, the solutions to the generalized
eigenvalue problem contain spurious states, also observed and
discussed in the original QSE paper~\cite{Colless:2018}. A solution of
the generalized eigenvalue problem exists only if $B$ is a
positive-definite matrix. Both $A$ and $B$ are Hermitian indefinite
matrices because they are created from noisy experimental data. The
positive-definiteness of $B$ is therefore not guaranteed. To
regularize the problem, we perform an eigenvalue decomposition of $B$,
select an optimal number of largest eigenvalues, and project both $A$
and $B$ onto the subspace corresponding to these eigenvalues. We then
solve the generalized eigenvalue problem with the projected
matrices. The number of largest eigenvalues of $B$ that are preserved
is maximized under the constraint that all of them are positive and
that the projected generalized eigenvalue problem contains no spurious
eigenvalues. The main problem is to decide if a particular eigenvalue
is a regular eigenvalue or a spurious eigenvalue. We were not able to
detect spurious eigenvalues by calculating the pseudospectra of the
generalized eigenvalue problem since the ground state seems to be too
sensitive to perturbations. Therefore, we detect spurious eigenvalues
by computing finite differences of the lowest eigenvalue as a function
of the number of preserved eigenvalues of $B$. We observe that the
lowest eigenvalue decreases monotonically until either $B$ becomes
indefinite or a spurious eigenvalue appears. The latter case results
in a sudden jump in the energy. If we selected all positive
eigenvalues instead of an optimized number of eigenvalues, spurious
energy levels would appear in the obtained potential energy curves.

All calculations were performed on the IBM Q Johannesburg quantum
computer. Since we used two qubits only, we estimated the overall
circuit fidelity from reported gate fidelities for all pair of qubits
and chose the pair with the highest overall fidelity. The selected
qubits were qubits Q0 and Q1.

\begin{figure}
  \centering
  \includegraphics{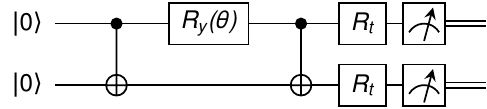}
  \caption{\label{circuit} A quantum circuit for the preparation of
    the UCC ansatz~\eqref{ansatz} and for the measurement of its
    expectation values. Gates $R_t$ perform a basis transformation
    that depends on the measured term.}
\end{figure}

The executed circuit is shown in Fig.~\ref{circuit}. It creates a UCC
ansatz~\eqref{ansatz} and performs a measurement in a selected
basis. Each $R_t$ gate is the $I$, $R_x(\pi/2)$, or $R_y(-\pi/2)$ gate
for measuring the corresponding qubit in the $Z$, $Y$, or $X$ basis,
respectively. Since the ansatz depends only on a single parameter, we
do not perform the VQE feedback loop to find the energy minimum. We
instead sample the full domain of $\theta$ and perform the
minimization on a classical computer later. In particular, we run the
circuit for 257 values of $\theta \in [-\pi, \pi]$ and for all nine
combinations of the $R_t$ gates. Each individual circuit was sampled
with 8192 shots. The raw data were unfolded~\cite{Nachman:2019} to
correct readout errors~~\cite{Kandala:2017, Dumitrescu:2018,
  YeterAydeniz:2019}. We then calculated all two-qubits expectation
values $\ev{P_1 P_2}_\theta$, where $P_i \in \{I, X, Y, Z\}$ is a
Pauli matrix acting on the $i$-th qubit. The same measured expectation
values were used for both the $\mathrm{H}_2$ and $\mathrm{Li}_2$
molecules.

Next, we performed a separate calculation for each molecule and for
each internuclear separation. The electronic wave functions were
represented using the cc-pVDZ basis set. The Hamiltonian terms were
calculated in OpenFermion~\cite{McClean:2020} using its interface to
Psi4~\cite{Parrish:2017}. The molecular Hamiltonian in the active
space with two orbitals was transformed using the Jordan--Wigner
transformation to a four-qubit Hamiltonian and then projected to a
reduced two-qubit Hamiltonian. We then calculated and smoothed the
expectation value of energy $E(\theta)$ from the measured expectation
values and found $\theta_\mathrm{min}$ that minimized it. The minimal
energy $E(\theta_\mathrm{min})$ is equivalent to the energy obtained
using the VQE algorithm. All measured expectation values $\ev{P_1
  P_2}_\theta$ were evaluated at $\theta_\mathrm{min}$ for their use
in later calculation stages.

We then created a list of expansion operators $O_i$. The operators
that changed the total spin number or that produced a state with the
norm below a cutoff when applied to the measured ground state were
removed from the list. The $A$ matrix was calculated by first
transforming operators $O_i H O_j$ using the Jordan--Wigner
transformation to four-qubit operators and subsequently projecting
them to two-qubit operators. The expectation value of each such
two-qubit operator was then estimated from the evaluated expectation
values $\ev{P_1 P_2}_{\theta_\mathrm{min}}$. The $B$ matrix was
created in a similar fashion.

\begin{figure}
  \centering
  \includegraphics{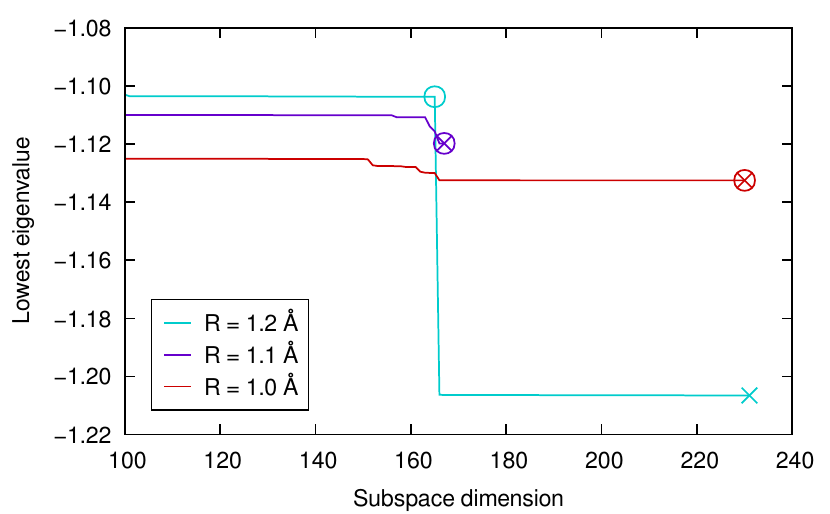}
  \caption{\label{regularization} Results of the eigenvalue
    regularization procedure for the $\mathrm{H}_2$ molecule and three
    values of its internuclear separation $R$. The plot shows the
    lowest eigenvalue of generalized eigenvalue
    problem~\eqref{eigenvalues} as a function of the number of largest
    eigenvalues of the $B$ matrix preserved in the regularization
    procedure. Crosses mark energies obtained when preserving all
    positive eigenvalues. Circles mark energies found using our
    regularization method.}
\end{figure}

Finally, we solved the generalized eigenvalue
problem~\eqref{eigenvalues} to obtain the ground state energy. The
solutions were sensitive to regularization because both matrices $A$
and $B$ were created from noisy data. We therefore first performed an
eigenvalue decomposition of $B$, selected an optimal number of largest
eigenvalues, and projected both $A$ and $B$ onto a subspace given by
the eigenvectors corresponding to the selected eigenvalues. The
generalized eigenvalue problem was solved using the projected $A$ and
$B$ matrices. Fig.~\ref{regularization} shows the idea behind the
regularization procedure. The lowest eigenvalue evolves in a
continuous way for internuclear separations $R = 1.0 \, \text{\AA}$
and $R = 1.1 \, \text{\AA}$, although the number of positive
eigenvalues of $B$ varies. For $R = 1.2 \, \text{\AA}$, a spurious
eigenvalue appears from 166 eigenvalues onward. The optimum is
therefore selected at 165 eigenvalues.

The calculations were performed both for zero virtual orbitals and the
selected number of virtual orbitals. The results for zero virtual
orbitals are equivalent to QSE results.

\begin{figure}
  \centering
  \includegraphics{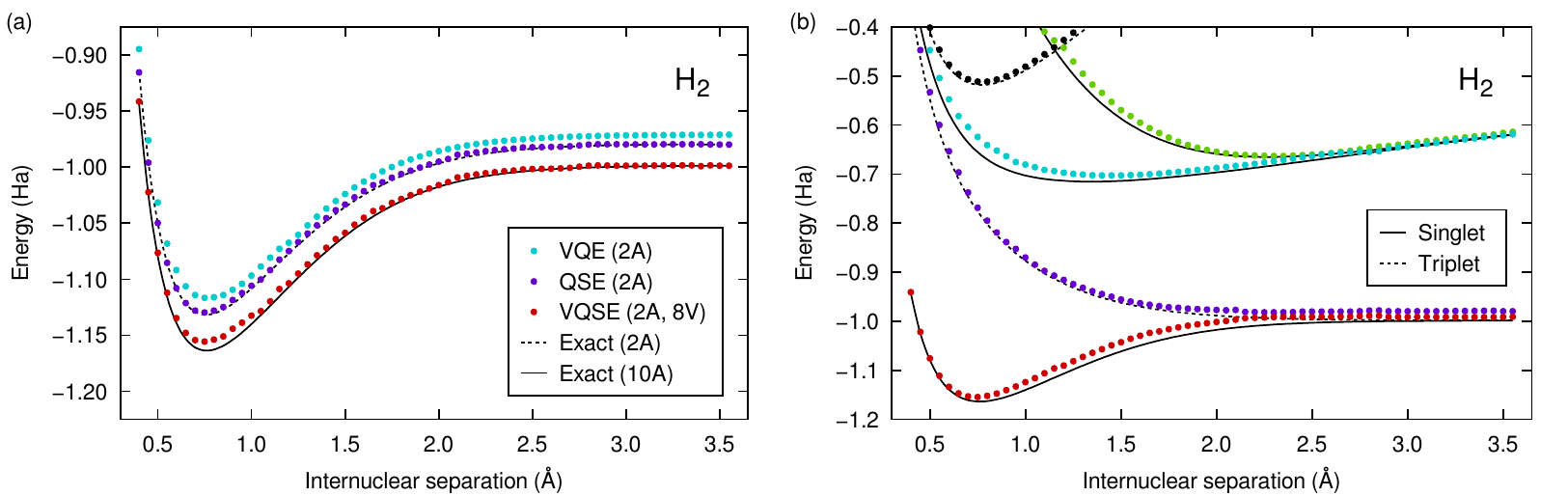}
  \caption{\label{hydrogen} Potential energy curves of the
    $\mathrm{H}_2$ molecule. (a) Ground-state energy obtained using
    several methods. Numbers and letters in parentheses indicate the
    number of active (A) and virtual (V) orbitals. (b) Five
    lowest-energy states calculated using the VQSE algorithm with two
    active and eight virtual orbitals. Solid and dashed curves
    represent the exact solutions in the combined space.}
\end{figure}

Potential energy curves for $\mathrm{H}_2$ in the ground state are
shown in Fig.~\hyperref[hydrogen]{\ref*{hydrogen}(a)}. The exact
solutions were found by full configuration interaction (FCI)
calculations within the chosen orbital subspace. Both VQE and QSE take
into account the active space only. QSE substantially improved the VQE
result. VQSE takes into account both the active and virtual
spaces. Executing VQSE with two active and eight virtual orbitals gave
rise to 296 expansion operators. VQSE significantly improved the
accuracy of the ground state energy. Its results with eight virtual
orbitals are close to the exact FCI result with ten virtual
orbitals. The largest difference is $9.9 \, \mathrm{mHa}$ at $R = 0.9
\, \text{\AA}$. The differences from the FCI solution are caused by
noise. We confirmed this by executing VQSE with data obtained from a
state-vector simulator. The noiseless result matched the FCI solution.

\begin{figure}
  \centering
  \includegraphics{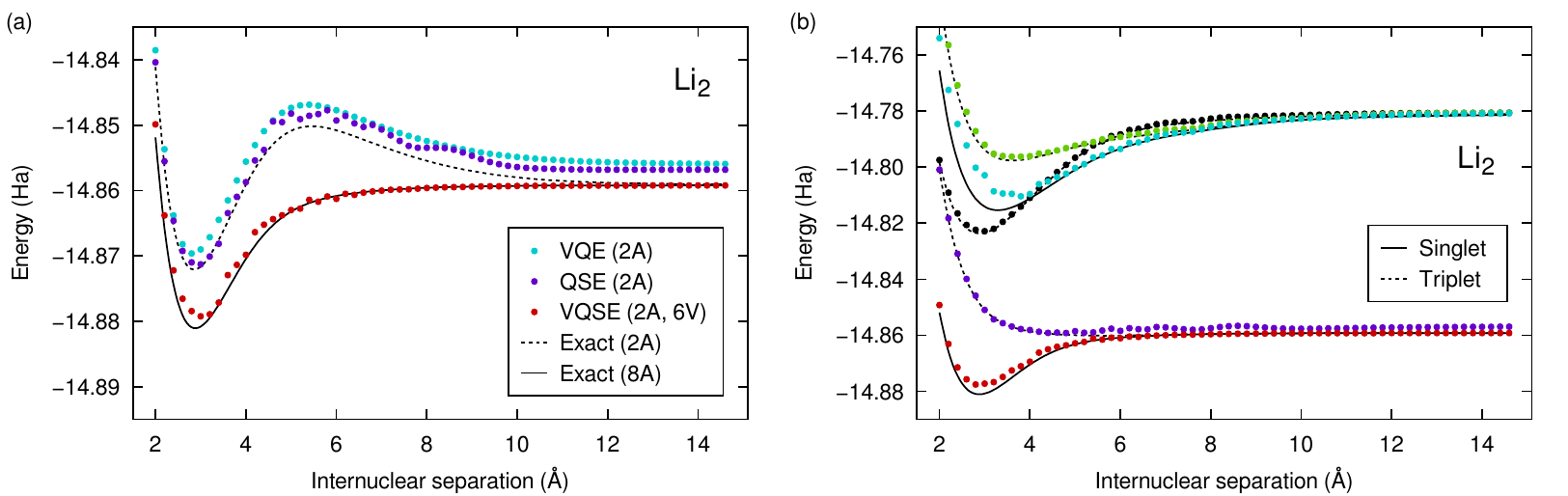}
  \caption{\label{lithium} Potential energy curves of the
    $\mathrm{Li}_2$ molecule. (a) Singlet ground-state energy obtained
    using several methods. Numbers and letters in parentheses indicate
    the number of active (A) and virtual (V) orbitals. (b) Five
    lowest-energy states calculated using the VQSE algorithm with two
    active and six virtual orbitals. Solid and dashed curves represent
    the exact solutions in the combined space.}
\end{figure}

Potential energy curves for $\mathrm{Li}_2$ in its singlet ground
state are shown in
Fig.~\hyperref[lithium]{\ref*{lithium}(a)}. Similarly to
$\mathrm{H}_2$, QSE improved the VQE result, but both methods take
into account orbitals in the active space only. Solutions in the
active space with two orbitals gave rise to an avoided crossing
manifesting itself as a hump in the energy at intermediate
internuclear separations. The hump is an artifact of a small active
space and disappears when one takes additional orbitals into
account. VQSE with two active and six virtual orbitals gave rise to
176 expansion operators. The VQSE method with two active and six
virtual orbitals produced a result close to the exact FCI solution
with eight active orbitals and without a hump. The largest difference
is $2.5 \, \mathrm{mHa}$ at $R = 2.6 \, \text{\AA}$. VQSE therefore
improved the potential energy curve both qualitatively and
quantitatively in this case.

A useful property of VQSE is its ability to find energies of excited
states. Lowest-energy states correspond to the smallest solutions of
the generalized eigenvalue problem. Our implementation finds both
singlet and triplet states due to the set of used expansion
operators~\eqref{operators}. The five lowest-energy states of both
molecules are shown in Fig.~\hyperref[hydrogen]{\ref*{hydrogen}(b)}
and Fig.~\hyperref[lithium]{\ref*{lithium}(b)}. For regularizing the
problem we did not use the same strategy as for the ground state since
the excited states do not follow the monotonicity property of the
ground state illustrated in Fig.~\ref{regularization}. Instead, we
removed spurious eigenvalues by preserving a constant number of the
largest eigenvalues of the $B$ matrix for all data points. In
particular, we preserved the 84 and 64 largest eigenvalues for
$\mathrm{H}_2$ and $\mathrm{Li}_2$, respectively. These were the
largest numbers of preserved eigenvalues that did not result in
spurious energy levels.

There are only two electrons in the active and virtual spaces of both
$\mathrm{H}_2$ and $\mathrm{Li}_2$ considered here. The combined
spaces can be covered by single and double excitations of the
Hartree--Fock states. VQSE therefore does not provide any advantage
over configuration interaction with singles and doubles (CISD) for
these problems. To show that VQSE can outperform CISD, we additionally
calculated the energy of the nitrogen dimer. Since it is challenging
to run this calculation on available quantum computers, we simulated
the VQSE algorithm classically.

\begin{figure}
  \centering
  \includegraphics{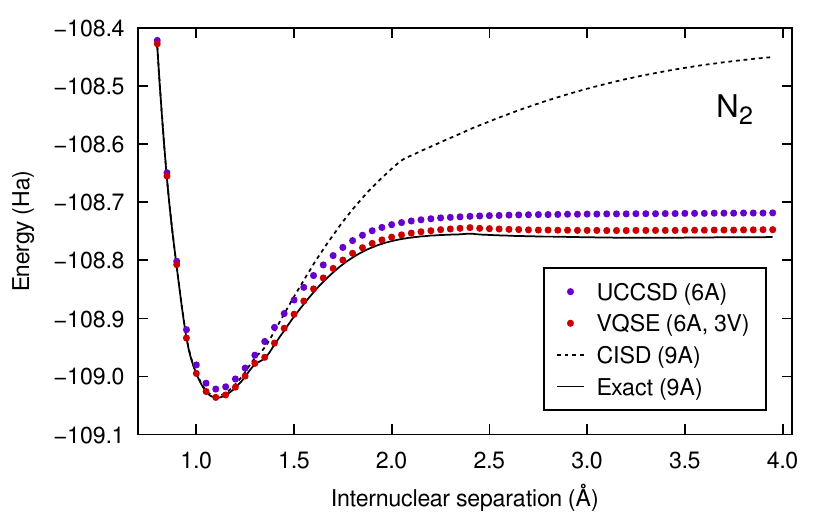}
  \caption{\label{nitrogen} Potential energy curves of the
    $\mathrm{N}_2$ molecule in its ground state obtained using a
    classical simulation of the VQSE algorithm. Numbers and letters in
    parentheses indicate the number of active (A) and virtual (V)
    orbitals.}
\end{figure}

We considered $\mathrm{N}_2$ in the cc-pVDZ basis with four core
orbitals, six active orbitals with six electrons, and three virtual
orbitals. The UCC ansatz with singles and doubles
(UCCSD)~\cite{Bartlett:1989, Taube:2006} was used as the ansatz for
the ground state in the active space. We included only excitation
operators preserving the $z$-component of the total spin. The
resulting ansatz had 117 parameters. We note that more compact
ansatzes, such as ADAPT-VQE~\cite{Grimsley:2019},
$k$-UpCCGSD~\cite{Lee:2019}, sequences of Jastrow-type
operators~\cite{Matsuzawa:2020}, or UCCSD variants~\cite{Sokolov:2020}
are possible but were not explored in this work. Both the Hamiltonian
and the ansatz were transformed to the qubit space using the
Jordan--Wigner transformation and the expectation value of the
Hamiltonian was calculated numerically. The ansatz wavefunction was
calculated exactly and without any approximations, such as the
first-order Suzuki--Trotter approximation, that would be required to
implement the ansatz on a quantum computer. The energy minimum was
found using the L-BFGS-B algorithm~\cite{Byrd:1995, Zhu:1997,
  Morales:2011}. We then directly calculated matrix elements of the
$A$ and $B$ matrices from the optimized ansatz and used VQSE to find a
correction to the ground-state energy. We compare the VQSE results
with those of FCI and CISD calculations obtained from Psi4. The
results are shown in Fig.~\ref{nitrogen}. The UCCSD VQE in the
six-orbital active space provides a significant improvement over CISD
for this problem. With the inclusion of additional virtuals through
the VQSE algorithm, we were able to obtain results that are close to
the FCI results in nine orbitals. The simulation was exact, so instead
of using our regularization procedure, we projected out space
corresponding to eigenvalues of the $B$ matrix close to zero. The
absence of spurious eigenvalues demonstrates that their source is
indeed the noise from the quantum computer. The optimized state in the
UCCSD form slightly differs from the exact ground state in the active
space, which explains the remaining error. We emphasize that even in
an ideal noiseless calculation presented here, the VQE algorithm could
not obtain the exact ground state energy. The ansatz with single and
double excitations is specified by a limited set of parameters and
therefore does not cover the full active space. We also observed that
the minimal energy depends significantly on the properties of the
classical optimizer and on the initial guess of the ansatz
parameters. We obtained the best results with all parameters being
initially zero.

Even though current quantum computers have many qubits, it is still a
major challenge to perform large-scale chemistry calculations on
them. For example, the UCCSD circuits for the $\mathrm{N}_2$ molecule
would require thousands of entangling quantum gates. The time required
to execute each circuit would be far beyond the coherence times of
existing qubits allowing for hundreds of gates at most. To our
knowledge, the largest chemistry calculation performed to date on a
quantum computer involved 12 qubits, 72 entangling gates, and 36
variational parameters~\cite{Arute:2020}. The authors studied the
Hartree--Fock state of a chain of 12 hydrogen atoms. Yet another
bottleneck is the classical optimization loop in the VQE
algorithm. Our classical simulation required 110,683 evaluations of
the Hamiltonian expectation value. If one evaluation took one minute
on a quantum computer, which is a very optimistic assumption, it would
take months for the algorithm to converge. Furthermore, even ideal
quantum computers produce statistical output containing shot
noise. Classical optimizers are sensitive to noise, which requires
careful selection of appropriate optimization algorithms and tuning
their parameters~\cite{McClean:2018, Lavrijsen:2020}. Overcoming these
challenges will require major advances in quantum hardware and
algorithms.

We implemented the VQSE method proposed in Ref.~\cite{Takeshita:2020}
on the IBM Q Johannesburg quantum computer and used it to calculate
the potential energy curves of the $\mathrm{H}_2$ and $\mathrm{Li}_2$
molecules. The noise of the quantum hardware is found to significantly
impact the classical generalized eigenvalue problem and we developed a
robust mathematical approach to address this issue. The obtained
results show a significant improvement in accuracy over the VQE and
QSE methods with the same number of qubits. In the present work, we
used only two qubits to model molecules with up to ten orbitals, which
would typically require twenty qubits using other algorithms. We also
demonstrated the effectiveness of VQSE for larger problems by finding
the ground-state potential energy curve of $\mathrm{N}_2$. VQSE is
therefore a promising method for studying chemical systems on
near-term quantum computers.

\begin{acknowledgement}

We thank Tyler Takeshita for helpful discussions. This work was
supported by the DOE under contract DE-AC02-05CH11231, through the
Office of Advanced Scientific Computing Research (ASCR) Quantum
Algorithms Team and Accelerated Research in Quantum Computing
programs. This research used resources of the Oak Ridge Leadership
Computing Facility, which is a DOE Office of Science User Facility
supported under Contract DE-AC05-00OR22725.

\end{acknowledgement}

\begin{tocentry}
  \centering
  \includegraphics[height=4.45cm]{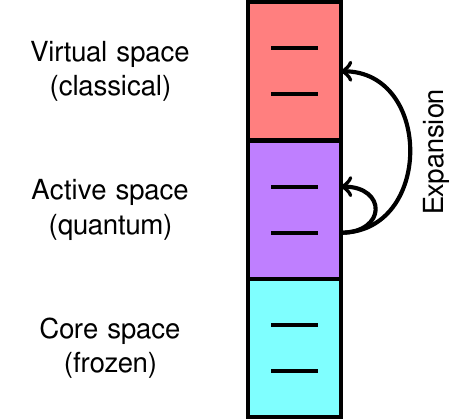}
\end{tocentry}

\bibliography{main}

\end{document}